\documentclass[12pt]{article}  
\usepackage[utf8]{inputenc}
\usepackage[english]{babel}
\usepackage[style=authoryear,backend=biber]{biblatex}
\usepackage[margin=1in]{geometry}
\usepackage{setspace}
\usepackage{times}  
\usepackage[T1]{fontenc}
\usepackage{graphicx}
\usepackage{caption}
\usepackage{booktabs}
\usepackage{csquotes}
\usepackage{float}
\usepackage{longtable}
\usepackage{multirow}
\usepackage{multicol}
\usepackage{tabularx}
\usepackage{xcolor}
\usepackage{ltablex}
\usepackage{xltabular}
\usepackage{tabularray}
\usepackage{amsmath, amssymb}
\usepackage{hyperref}
\usepackage{algorithm}
\usepackage{algorithmic}
\usepackage{verbatim}
\usepackage{listings}
\newcommand{\ignore}[1]{}
\newtheorem{definition}{Definition}
\newtheorem{insight}{Insight}

\renewcommand{\cite}{\parencite}
\floatstyle{ruled}
\DeclareCaptionStyle{ruled}{labelfont=normalfont,labelsep=colon,strut=off}

\title{Towards Proactive Defense Against Cyber Cognitive Attacks}
\author{Bonnie Rushing, Mac-Rufus Umeokolo, Dr. Shouhuai Xu}
\date{\today}

\begin{document}

\maketitle

\section*{Abstract}
\noindent
Cyber cognitive attacks leverage disruptive innovations (DIs) to exploit psychological biases and manipulate decision-making processes. Emerging technologies, such as AI-driven disinformation and synthetic media, have accelerated the scale and sophistication of these threats. Prior studies primarily categorize current cognitive attack tactics, lacking predictive mechanisms to anticipate future DIs and their malicious use in cognitive attacks. This paper addresses these gaps by introducing a novel predictive methodology for forecasting the emergence of DIs and their malicious uses in cognitive attacks. We identify trends in adversarial tactics and propose proactive defense strategies.

\noindent Keywords: Cyber cognitive attacks, Disruptive innovations, AI-driven disinformation, Psychological biases, Information warfare

\textit{Disclaimer: “The views expressed are those of the author and do not reflect the official policy or position of the US Air Force Academy, US Air Force, Department of Defense, or the US Government.”}

\section{Introduction}

Cyber cognitive attacks can utilize advanced digital tools to influence or manipulate their decision-making processes. These attacks fall under the intersection of information operations and cyber social engineering attacks. They target the subconscious by exploiting and managing mental biases, shortcuts, or reflexive thinking to elicit a person's immediate reaction and perception of reality \cite{blacksea}. They affect targets' sensations, attention, memory, and mental operations, causing humans to be misled into fallacious reasoning and manipulated decisions \cite{Linan}. They provoke thought distortions and influence decision-making, adversely affecting individuals or groups who may be more willing to believe like-minded claims without sufficient evidence and make hasty, irrational choices \cite{NATOscience, Crupi_Mejova_Tizzani_Paolotti_Panisson_2022}, and reshaping individuals’ deep underlying knowledge to manipulate their subconscious perceptions \cite{ox}. 

The situation is getting worse because cyber cognitive attackers have been abusing the digital revolutions in information technology to amplify their Tactics, Techniques, and Procedures, or TTPs \cite{mitre}, where \textit{Tactics} represent an attacker's tactical goal (i.e., the reason for performing a malicious action),  \textit{Techniques} represent how an attacker performs an action to achieve its tactical goal, and \textit{Procedures} represent specific implementations of attack techniques.
In particular, the proliferation of Disruptive Innovations (DIs) in technology, such as AI-enabled bots and synthetic media, has amplified these attacks, enabling adversaries to execute precise, large-scale cognitive attacks with unprecedented speed and sophistication. The state of the art reveals that traditional defenses are ineffective against cognitive attacks leveraging DIs, leaving critical gaps in security measures \cite{blacksea}. This calls for new approaches to countering cognitive attacks exploiting DIs, which motivates the present study. Our methodology aims to empower populations by addressing cyber challenges and public concerns, such as disinformation, thereby sustaining overall progress in democratic and socioeconomic development through proactive defenses \cite{lin2023mapping}. 

\smallskip

\noindent{\bf Our Contributions}. This paper makes two contributions.
First, we initiate the study on the following question: \textit{Is it possible to predict the emergence of DIs and proactively design countermeasures to defend against cognitive attacks that exploit them?} To tackle the problem, we propose a methodology that consists of two key components: (i) predicting the emergence of DIs based on historical data by leveraging both {\em quantitative} models (for predicting {\em when} DIs may emerge in the future) and {\em qualitative} models (for predicting {\em what} attributes new DIs may possess); (ii) leveraging the predicted DIs to guide the design of proactive defenses against cognitive attacks that exploit these predicted DIs.
It is worth mentioning that the present study is multidisciplinary in nature because DI is a concept in management science \cite{ANTONIO2023122274}, which, to the best of our knowledge, is introduced to the cybersecurity domain as a model to predict emerging DI tools for the first time.

Second, we demonstrate the feasibility and usefulness of the framework by conducting a case study, which leads to 11 DIs that emerged between 1971 and 2017, where 2017 is chosen to validate the predictions occurring during 2018-2024. Since the dataset only consists of 11 data points, which may be too small to consider advanced models, we use simple prediction models. We show these simple models can indeed predict the emergence of cognitive attacks that exploit, for instance, the humanoid robots DI invented in 2021, and how predictions can guide the design of proactive defenses against cognitive attacks that exploit DIs. 

The case study also leads to several findings, such as the following. 
(i) Methodology-wise, integrating qualitative prediction and quantitative prediction is both feasible and useful in the study of cyber cognitive attacks.
(ii) Mapping historical DIs to existing frameworks (MITRE ATT\&CK and DISARM) reinforces the importance of structured analysis. This mapping allows for identifying TTPs associated with each DI, enabling the development of targeted countermeasures.
(iii) Applying qualitative attributes (Model 2) provide a comprehensive method for assessing potential DIs by scoring their disruptive impact across key dimensions, such as communication pathways, data collection, victim targeting, and social networks. This qualitative understanding complements the quantitative characterization by offering contextual insights into predicted DIs.
(iv) The societal and technological implications of DIs need more research.

\ignore{

\subsection{Contributions}
This paper makes the following novel contributions:

\begin{itemize}
    \item \textit{Integration of Established Frameworks}: We integrate MITRE ATT\&CK and DISARM frameworks to categorize DIs and their role in cognitive attacks, providing a structured approach to analyzing adversary tactics.
    \item \textit{Development of Predictive Models}: We introduce two predictive models—Time Span Mean and Linear Regression Average—that forecast the emergence of future DIs based on historical trends.
    \item \textit{Proactive Defense Framework}: We propose a set of defense mechanisms, including prediction-based threat simulations, education, and incident response planning, designed to mitigate emerging cognitive threats.
    \item \textit{Actionable Metrics for Prediction}: We provide clear metrics and a step-by-step process for applying our predictive models to real-world cybersecurity contexts, enabling proactive risk management.
    \item \textit{Interdisciplinary Insights}: Our work bridges the gap between prediction modeling and cybersecurity defense strategies, offering valuable insights for researchers, policymakers, and practitioners across multiple fields.
\end{itemize}

}

\smallskip

\noindent{\bf Related Work}. We divide related prior studies into three categories: cyber social engineering attacks, emerging technologies, and predictive cybersecurity analytics. 

Cyber social engineering attacks are broader than cyber cognitive attacks, with a common feature of exploiting humans' psychological factors, especially cognitive psychological factors \cite{longtchi2024internet}. There is a recent trend in studying cyber social engineering attacks through psychological lenses. However, cyber cognitive attacks have not been systematically investigated from a cognitive psychology perspective, which complements our study. 

Emerging technologies, especially DIs (e.g., algorithmic targeting), are transformative and can reshape the threat landscape, including cognitive attacks. Existing studies \cite{ox, bots} primarily focus on retrospective analyses of technological disruptions. By contrast, we investigate how to predict the emergence of DIs and leverage the predictions to guide the design of proactive defenses against the anticipated cognitive attacks that exploit those DIs.

Predictive cybersecurity analytics can forecast cyber threats \cite{longtchi2024internet} but often require large enough datasets to build advanced models. These advanced modeling approaches would not apply to our setting because the DI dataset is too small. This explains why we use simple predictive models, which are nevertheless useful.

\smallskip

\noindent{\bf Paper Outline}. Section \ref{sec:attack-and-DI} discusses cyber cognitive attacks and DIs. Section \ref{sec:Methodology} presents our research methodology. Section \ref{sec:case-study} describes our case study. Section \ref{sec:discussion} discusses the limitations of the present study. Section \ref{sec:Conclusion} concludes the paper.

\section{Cognitive Attacks and DIs}
\label{sec:Technologies}
\label{sec:attack-and-DI}

Non-cyber cognitive attacks existed prior to the cyber era, such as those documented since China’s Warring States period (475–221 BCE), with a strategic landscape marked by deceptive competition. Historical cognitive attack objectives include undermining hegemonic powers, fostering complacency and discord, influencing governments and voters, promoting selective narratives, and shaping perceptions and decision-making \cite{ox}. 

In the cyber era, successful cyber cognitive attacks require potential victims (i.e., Internet or cyberspace users) to encounter and consume an attacker's malicious content, such as disinformation \cite{ieuniversity}. Although cyber cognitive attackers continue to use old deception methods, they have been abusing DIs to amplify their effects simply because DIs can increase their effectiveness, proliferation, and international security threats \cite{ox}. For instance, one DI could enable a cyber cognitive attacker to reach its international audience in near real-time. 

To thwart cyber cognitive attacks that may or may not exploit DIs, we must adequately characterize them. For this purpose, we propose leveraging the DISARM framework \cite{DISARM}, which was introduced to describe disinformation incidents, and the MITRE ATT\&CK framework \cite{mitre}, which primarily describes cyber attacks against IT networks but can somewhat describe cognitive attacks and cyber social engineering attacks. These frameworks describe attacks--cyber, cyber social engineering, and cognitive alike--in terms of TTPs (Tactics, Techniques, and Procedures). 
For example, DISARM contains a technique known as ``{\em T0086.002 - Develop AI-Generated Images (Deepfakes)}'' under its tactic TA06, which describes the use of deepfakes to fabricate images, videos, or audio \cite{DISARM}. This is important because the mapping allows for a structured and systematic analysis of how DIs may be exploited by cyber cognitive attacks, which could guide the design of countermeasures. 
In summary, we focus on:

\begin{definition}
[DI-enabled cyber cognitive attacks]
\label{def:DI-enabled cyber congitive attacks}
These are the cyber cognitive attacks that exploit DIs that introduce new, expand existing, or automate methods for communications, data collection, social networking, or targeting in their TTPs to amplify their effect.
\end{definition}

Definition \ref{def:DI-enabled cyber congitive attacks} says that we treat technology or innovation as a DI if it can impact cyber cognitive attacks, namely that the technology can enable disruptive changes to cognitive attacks. This means standard technology upgrades, concept improvements, speed increases, and hardware updates are not treated as DIs. 
Definition \ref{def:DI-enabled cyber congitive attacks} also implies that although the nature of cognitive attack objectives remains largely the same throughout history, cyber cognitive attacks have been advanced substantially in terms of their {\em speed} and {\em scale} because of their exploitation of DIs. 

\section{Methodology}
\label{sec:Methodology}

Our methodology has four steps: {\em DI dataset collection}, {\em data preprocessing}, {\em modeling and prediction}, and {\em designing proactive defense}, which are elaborated below.

\noindent{\bf DI Dataset Collection}.
This step identifies and collects a dataset of historic DIs pertaining to cognitive attacks as shown in Definition \ref{def:DI-enabled cyber congitive attacks}. In terms of scope, we advocate focusing on the technologies that can alter how cognitive attackers operate, starting with the first cyber communication tools (e.g., email and direct messaging) while disregarding pre-cyber technologies such as radios and television. 

\ignore{
\begin{figure}[H]
    \centering
    \includegraphics[width=0.75\linewidth]{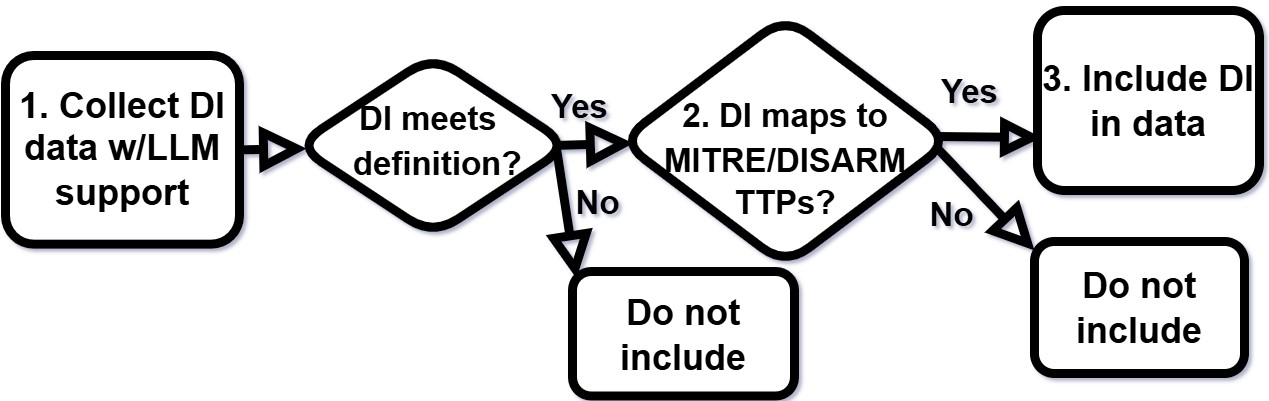}
    \caption{Procedure for preparing DI dataset}
    \label{fig:inclusionchart}
\end{figure}

Figure \ref{fig:inclusionchart}}

The DI data collection procedure can be understood as follows. (i) Collect frequent DI trends, with the support of a Large Language Model (LLM) like ChatGPT AI, to deal with the large number of source materials, which include expert opinions such as those documented in Elon University's 2016 survey \cite{elon}. 
(ii) Filter the identified DIs via Definition \ref{def:DI-enabled cyber congitive attacks} to assure relevance. Domain experts in cyber cognitive attacks often manually perform this final filtering process.
(iii) Map the DIs that survive the filtering process to the TTPs of DISARM \cite{DISARM} and MITRE ATT\&CK \cite{mitre} frameworks. This mapping process is often done manually but could be automated, which requires further study in the future. For instance, the DI known as Virtual Reality can be mapped to the DISARM  tactic known as ``{\em Virtual Engagement Manipulation}" and the MITRE ATT\&CK technique known as ``{\em Establish Accounts}.'' to create personas during targeting.

\noindent{\bf Data Preprocessing}.
This step organizes the selected DIs in a data structure to support the subsequent modeling and prediction step. For each DI, the essential attributes include:
(i) {\em DI Name}: This identifies a DI.
(ii) {\em DI Year}: This is the year the DI was publicly introduced or employed.
(iii) {\em DI Inclusion Rationale}: This is the justification for including the DI, including its disruptive impact attributes.
(iv) {\em DI Description}: This is a brief description of the DI’s evolution or application since its initial implementation, if applicable.
(v) {\em DI Mapping}: This describes the TTPs to which the DI is mapped. The mapping can be done manually but should ideally be done with AI assistance for large-scale modeling.

\noindent{\bf Modeling and Prediction}.
We propose using appropriate predictive models, including statistical and qualitative models, to forecast the evolution of cyber cognitive attacks that may exploit future DIs. Ideally, these models should be able to predict the year of a future DI emergence and its attributes; the latter can be leveraged to design proactive defenses against the cognitive attacks that exploit the new DIs.

\noindent{\bf Designing Proactive Defense}.
This step is to leverage the predicted DIs that may be exploited by future cognitive attacks to proactively design countermeasures against them. Given that cyber cognitive attacks are at the intersection of cybersecurity (or computer science) and cognitive science, effective countermeasures must go beyond traditional solutions. This observation prompts us to advocate considering a range of stakeholders, including cybersecurity teams in both public and private sectors, government regulatory agencies, academic and industry researchers, technology developers, public awareness campaign organizations, and international organizations (e.g., the United Nations). These interdisciplinary leaders should execute mitigation TTPs related to cognitive attacks to benefit society.

\section{Case Study}
\label{sec:case-study}
To demonstrate the usefulness of the methodology, we apply it to conduct the following case study. 
\label{sec:framework}


\subsection{Select and refine cognitive DIs} 
Under the guidance of the methodology, we consider technologies dating back to the 1970s when email and directed messages were introduced. We leverage the 1,537 expert responses in the Elon University survey \cite{elon} and use them as input to ChatGPT 4's LLM to identify frequent text span trends via the following prompt: \textit{What unique cognitive attack disruptive innovations (DI), including technologies (e.g., social media) and innovations (e.g., troll factories), are mentioned by participants in this survey? What are the cyber cognitive attack attributes (e.g., introducing cyber communications, social network accounts, and targeting) of each DI?}

Then, we manually filter the DIs returned by ChatGPT 4's AI output to identify the true DIs that could be exploited to wage cognitive attacks and meet our definition of DI. Additionally, through this methodology, we developed the following cyber cognitive attack DI attributes.
(i) Cyber Communication Pathways, which we describe as a required venue(s) for attackers to contact victims online \cite{elon}.
(ii) Data Collection, which we describe as attackers' capabilities to gather victims' information from Internet-based sources \cite{trackers}.
(iii) Victim Targeting, which we describe as attackers' capabilities to predict online users' preferences based on data analysis of victims, filtering content related to a target \cite{breese1998empirical}.
(iv) Social Networks, which we describe as platforms that enable users to create personal profiles on the internet, facilitating connections, interactions, and content sharing online. Social networks can also be exploited by attackers to identify and contact potential victims \cite{socialmedia}.

This leads to the following DIs and the numbers of their mentions in the survey \cite{elon} and inclusion rationale based on cognitive attack DI attributes:
\begin{enumerate}
\item E-mail/direct messaging: mentioned 26 times. 
This technology, introduced in 1971 \cite{motherboard2010disinformation}, is disruptive because it introduced cyber communications that made cyber cognitive attacks possible (e.g., sending precise messages based on attackers' advanced understanding of victims \cite{ieuniversity}).

\item Online gaming platforms: mentioned 18 times. This technology, introduced in 1980, was born when outside users connected through the ARPANET \cite{mud}. Since then, online gaming has expanded massively, enabling users to play and chat with other users worldwide. It is disruptive because it expanded pathways to communicate and conduct cyber cognitive attacks.

\item Virtual reality (or augmented/mixed reality): mentioned 9 times. This technology, introduced in 1991, enhances gaming platforms and other applications by providing advanced online experiences with 3D visuals, with further developments including hand scanning and eye tracking \cite{VR}. It is disruptive because it expanded communication pathways for cyber cognitive attackers to reach targets.

\item Cookie technology for user tracking and data collection: mentioned 15 times.
This technology, introduced in 1994, enables automated data collection in an HTTP environment, and it is difficult for web users to restrict what data can be collected by cookies \cite{trackers}. 
The collected data could include personal data and preferences and thus could be abused to enable targeted cyber cognitive attacks with contextually relevant content to maximize visibility and impact \cite{trackers2}. This technology is disruptive because it can automatically collect data from a massive population of users, and adversaries could abuse the collected data to wage cognitive attacks against victims.

\item Algorithmic targeting: mentioned 39 times. This technology was introduced in 1995 to use collaborative filtering to predict online users' movie preferences based on their votes on movies they liked  \cite{breese1998empirical}. Since then, data analysis and algorithms have increasingly impacted the information presented to users, with each person seeing a personalized version of reality as demonstrated by Facebook’s news feed, Twitter’s (X) Timeline, Google's ranking system, and Netflix's and YouTube’s recommender system \cite{bots}. 
Attackers can abuse algorithmic targeting to wage cognitive attacks against victims based on their known fears and beliefs \cite{ieuniversity}. This technology is disruptive because it can automatically target a massive population of victims with personalized attacks, possibly by leveraging AI/ML technologies.

\item Social media: mentioned 227 times. This technology was introduced in 1997 as manifested by the first social networking website \url{SixDegrees.com}. It typically allows members to create personal profiles, maintain friends lists, and contact one another through private messaging systems \cite{socialmedia}. It is disruptive because it introduced social network accounts, expanded access to personal data, and expanded communication pathways, effectively widening venues for cognitive attackers to reach victims.

\item Mobile apps (applications): mentioned 17 times.
This technology was introduced in 1997 when the Nokia 6110 included a built-in version of the basic arcade game ``Snake,” which many consider the first mobile application (app).  Since then, the quantity and utility of apps have exploded exponentially, providing users easy access to programs that include social media, gaming, messaging, e-mail, personal data, healthcare, banking, shopping, and many more. It has been abused to wage cyber cognitive attacks \cite{apps}. It is disruptive because it has widely expanded access to personal data and communication pathways that can be abused to wage cyber cognitive attacks.

\item Bots (including social bots and software robots): mentioned 32 times. This technology was introduced in 1999 to allow user accounts on social media platforms to operate fully- or semi-automatedly. Their communications are designed to imitate human behaviors, ranging from automating individual elements of communication processes (e.g., liking or sharing), partially human-steered accounts with automated elements ('hybrid bots' or ‘cyborgs’), to autonomous agents equipped with AI/ML skills (e.g., Microsoft’s Zo1 or \url{Replika.ai} \cite{bots}). This technology is disruptive because it can be abused to automate social network accounts to wage cyber cognitive attacks at a massive scale.

\item IoT (Internet of Things, including wearable technologies): mentioned 9 times. This technology was introduced in 2007, as evidenced by Apple's iPhone, which began a new era of smartphones. By 2008, the number of connected IoT devices overtook the number of people worldwide, with exploding IoT devices in developed nations \cite{iot}. This technology is disruptive because malicious actors can use these devices to track and collect user data globally.

\item Trolls (troll factories): mentioned 69 times.
This DI emerged in media reports in 2015. It consists of organized group tactics that deliberately provoke and disrupt online discussions, often for political or ideological purposes. Troll factories may be government-sponsored and employ hundreds of attackers to achieve their goals using techniques such as fake news and hate speech. Trolls create fake identities and run their profiles on social media, creating an impression of account authenticity \cite{trolls}. It is disruptive because it can abuse social network accounts and communication pathways to wage cyber cognitive attackers.

\item AI synthetic media (or altered/fictional/fake content): mentioned 9 times. This technology was introduced in 2017 when actors use Large Language Models (LLMs) to generate contextually relevant, persuasive, and tailored disinformation at a pace that outstrips the ability of human moderators and existing automated systems to counteract effectively (e.g., ``DeepFake" pictures, modified voice, text, and videos \cite{socialmedia}). An exploration of the weaponization of AI in automating the spread of disinformation and the challenges in combating such threats can be found in \cite{CSET}. This is disruptive because AI synthetic media can generate high-quality multimedia content that can be abused to wage cognitive attacks.
\end{enumerate}

Note that the preceding 11 DIs were extracted from a 2016 survey \cite{elon}, before the introduction of the AI synthetic media DI, which is included because its relevance is documented \cite{CSET}.

\subsection{Map selected DIs to TTPs} 

Table \ref{Tab:mapping} summarizes the mapping of the selected 11 DIs to the TTPs in the MITRE ATT\&CK \cite{mitre} or DISARM \cite{DISARM} framework and the rationales for selecting the 11 technologies as DIs. To demonstrate how we obtained the TTPs, in what follows, we use mobile applications (apps) DI as one example.

\begin{table*}[h]
\caption{Rationale for selecting the 11 DIs and their mappings to DISARM and MITRE ATT\&CK TTPs}
 \centering
    \renewcommand{\arraystretch}{0.1} 
    \renewcommand{\arraystretch}{1.0} 
    \setlength{\tabcolsep}{2pt} 
    {\singlespacing 
    \scriptsize 

\begin{tabular}{|p{1.3cm}|p{2.5cm}|p{5.5cm}|p{6.8cm}|}
\hline
\textbf{DI} & \textbf{Year/ \textit{DI Attributes}} & \textbf{Expert Survey Example} & \textbf{MITRE ATT\&CK/DISARM TTPs} \\
\hline
\textbf{1. E-mail / Message Platforms} & 
1971 - Introduced \textit{cyber comm pathways} & 
Paul Edwards, University of Michigan: ``Social media will continue to generate increasingly contentious, angry, mob-like behavior... consistently observed since the early days of email.'' & 
DISARM: Technique: Email (Delivering content and narratives via email);
ATT\&CK: Technique: Spearphishing Attachment (social engineering attack against specific targets... attach a file to the spearphishing email) \\
\hline
\textbf{2. Gaming Platforms} & 
1980 - Expanded \textit{cyber comm pathways}  & 
Bryan Alexander Consulting: ``The number of venues will rise... providing cultural and psychological backing for abusive expression.'' & 
DISARM: Technique: Video Livestream (real-time comms); DISARM: Technique: Chat apps (messaging often automated with new delivery methods (e.g., online games)\\
\hline
\textbf{3. Virtual Reality} & 
1991 - Expanded \textit{cyber comm pathways}& 
Bryan Alexander: ``The expansion of virtual and mixed reality venues will increase the opportunities for manipulation.'' & 
DISARM: Technique: Build Network (operators build their own network, creating links between accounts... promote narratives and encourage further growth); ATT\&CK: Technique: Establish accounts\\
\hline
\textbf{4. Tracking and Data Collection} & 
1994 - Introduced automated \textit{data collection} & 
Randy Albelda, University of Massachusetts: ``How data collected from the Internet is used... reshapes how we think.'' & 
DISARM: Technique: Segment Audiences (create audience segmentation by features of interest... e.g., political affiliation, location, demographics); ATT\&CK: Technique: Location Tracking (track a device's physical location) \\
\hline
\textbf{5. Algorithm Targeting} & 
1995 - Introduced automated \textit{victim targeting} & 
Karen Blackmore, University of Newcastle: ``Algorithms are degrading our ability to debate by reinforcing views.'' & 
DISARM: Technique: Manipulate Platform Algorithm (increase content exposure, avoid content removal...) DISARM: Technique: Leverage Echo Chambers/Filter Bubbles (filter bubble refers to an algorithm's placement of an individual in content that they agree with...)\\
\hline
\textbf{6. Social Media} & 
1997 - Introduced\textit{ social network} accounts for \textit{cyber communications pathways} & 
Vint Cerf, Google: ``Social media bring every bad event to our attention... leading to an overall sense of unease.'' & 
DISARM: Technique: Social Networks (expression through virtual communities and networks); DISARM: Technique: Post inauthentic social media comment\\
\hline
\textbf{7. Apps} & 
1997 - Expanded access to \textit{data collection} \& \textit{comm pathways} & 
Matt Bates: ``Every digital application can be used for surveillance and control.'' & 
DISARM: Technique: Dating Apps (broadcast capability that allows for real-time communications); ATT\&CK: Technique: Stored Application Data (access device app data) \\
\hline
\textbf{8. Bots} & 
1999 - Automated\textit{ social network} accounts & 
Georgia Tech Professor: ``The looming influence of bots... manipulates opinions.'' & 
DISARM: Technique: Post inauthentic social media comment; DISARM: Technique: Acquire Botnets (group of bots that can function in coordination with each other) \\
\hline
\textbf{9. IoT} & 
2007 - Expanded \textit{cyber comm pathways} \& \textit{data collection} & 
Thorlaug Agustsdottir: ``The Internet of Things will change our use of everyday technology...'' & 
DISARM: Tactic: Amplification; ATT\&CK: Technique: Transmitted Data Manipulation (adversaries may intercept and change transmitted data)\\
\hline
\textbf{10. Troll Factories} & 
2015 - Advanced \textit{social network} accounts \& \textit{comm pathways} & 
Laurent Schüpbach: ``Russian troll armies are a good example of organized influence.'' & 
DISARM: Technique: Trolls amplify and manipulate; DISARM: Technique: Conduct Swarming (coordinated use of accounts to overwhelm information space) \\
\hline
\textbf{11. AI Synthetic Media} & 
2017 - Advanced \textit{cyber comm pathways}' content & 
Pew Research Center: ``Efforts to influence through misleading stories, photos, and videos.'' & 
DISARM: Technique: Develop AI-Generated Text (synthetic text composed by computers); DISARM: Technique: Develop AI-Generated Audio (falsified photos, videos, sound) \\

\hline
\end{tabular}
\label{Tab:mapping}} 
\end{table*}

For this example, we identified a DISARM tactic through their website's framework explorer, ``dating apps," within the framework where it is described as providing adversaries with ``broadcast capability that allows for real-time communication." Additionally, MITRE ATT\&CK listed ``deliver malicious app" as technique ID T1476 on their website, describing how it may be employed for cognitive attacks. In this case, MITRE explained that malicious apps may be included with spearphishing links, email, and text messages. 

\subsection{Data Preprocessing}

Table \ref{Tab:mapping} highlights the preprocessed dataset. It contains 11 DIs (i.e., rows). Each DI is described via the following attributes in the columns in the following manner: DI name, DI year, DI inclusion rationale (attributes), DI description (expert survey example), and DI mapping (to TTPs).

\label{sec:Findings}

\subsubsection{Model Fitting and Prediction}

Let $n$ denote the number of DIs and $n=11$ in our dataset. Let $X_i$ denote the year during which the $i$th DI was introduced (e.g., $X_1=1971$ for the e-mail DI) where $1\leq i \leq n=11$ for the dataset. Note that in general, $i< j$ implies $X_i \leq X_j$ as multiple DIs may occur in the same year, also as shown in our dataset.

Our case study employs two predictive models: Model 1, which predicts when future DIs may emerge, and Model 2, which predicts the attributes of future DIs to inform the design of proactive defense. 

{\bf Model 1: Prediction based on Linear Regression}.

We used the dataset to fit a standard linear regression model $Y=a_0 + a_1 X$ via the Python library, where $X$ represents the year during which a DI was introduced (e.g., $X_1=1971$ and $X_2=1980$ as mentioned above)
and $Y$ represents the cumulative number of DIs that have been invented since 1971 and up to year $X$ (e.g., $Y_1=1$).  

The fitted model is $Y=-460.4953271028037 + 0.23364486 X$.

\begin{figure}[!htbp]
    \centering
\includegraphics[width=.7\linewidth]{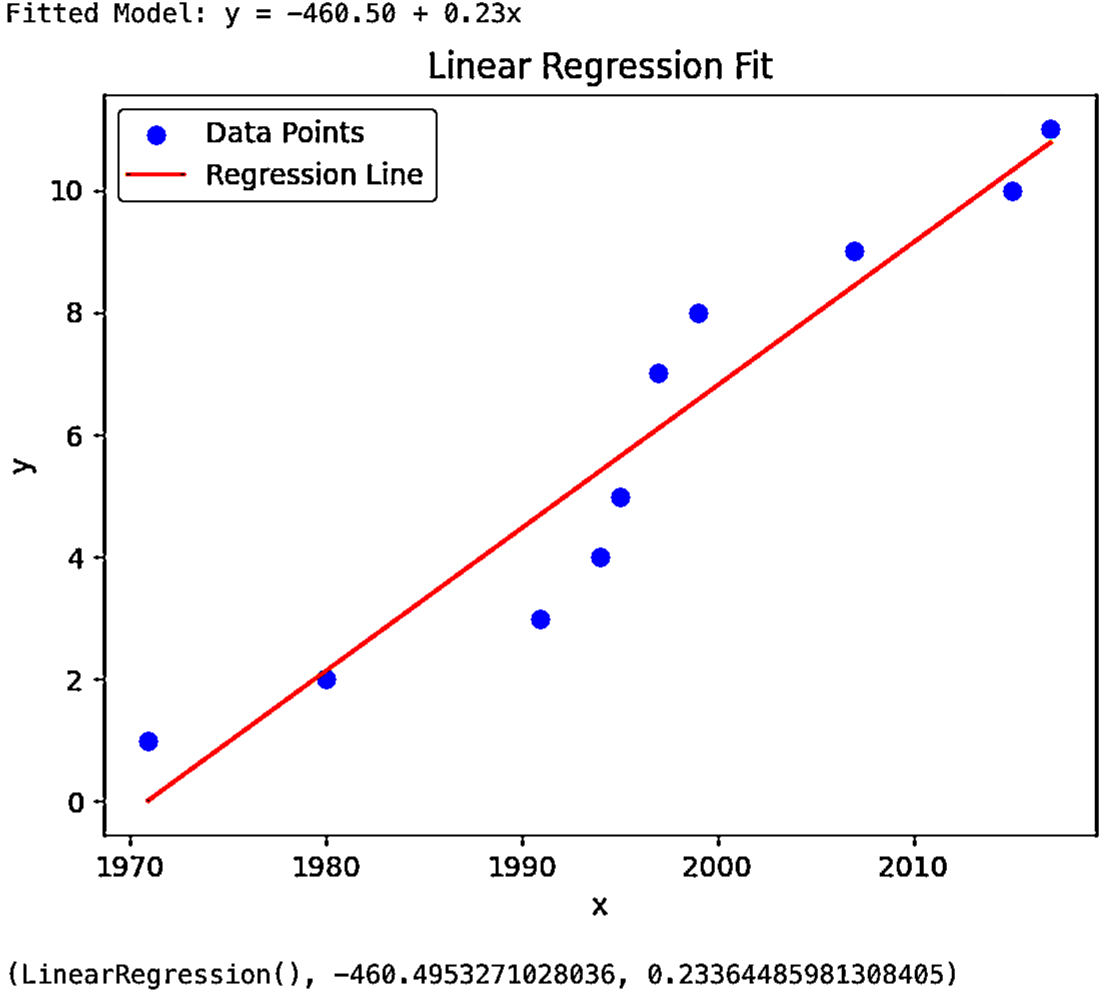}
    \caption{Linear regression of DI emergence timeline}
    \label{fig:model2}
\end{figure}

Figure \ref{fig:model2} plots the fitting result. Based on the fitted model, we set $Y=12$ to derive $X=2,022.28$, meaning that the 12th DI was expected to emerge in 2022.

{\bf Model 2: Prediction based on Qualitative DI Attributes}.
We use qualitative analysis to predict DI attributes 
(e.g., introducing, expanding, advancing, or automating cyber \textit{communication pathways, data collection, victim targeting, and social networks}) mentioned above.

Investigators score each attribute for a suspected emerging DI. For example, if researchers believe that a new DI has a ``high" impact for any attribute, they would record a score of 3. For simplicity, the scale we use is 1 (Low impact) to 3 (High impact), and we define a threshold of 9 or higher. If the total score of all attributes is 9 or higher, we flag the suspected DI for consideration for inclusion into the data if it meets all other requirements in the methodology. To demonstrate the usefulness of model 2's approach, we use \textit{AI Synthetic Media} as a case study in Table \ref{tab:model3}. Based on research and publications \cite{socialmedia}, we scored this DI as follows:

\begin{table}[htpb!]
\centering
\begin{tabular}{|l|l|} \hline  
\textit{Cyber Cognitive Attack DI Attributes}  &\textit{Score (1-3)}\\ \hline 
Cyber Communication Pathways  &3\\ \hline 
Data Collection  &1\\ \hline 
Victim Targeting  &2\\ \hline 
Social Networks  &3\\ \hline
\textit{Total score $(S)$}&\textit{9}\\\hline

\end{tabular}
\caption{Model 2 Scoring Example: AI Synthetic Media}
\label{tab:model3}
\end{table}

\textit{Cyber Communication Pathways: 3, high impact}; Deepfakes and other altered content can generate persuasive multimedia, automating the spread of disinformation across any communication platform.
\textit{Data Collection: 1, low impact}; while the multimedia does not directly collect data, it may be used with other DIs that can.
\textit{Victim Targeting: 2. medium impact}; when cognitive attackers employ LLMs to generate contextually relevant and tailored disinformation, victim targeting becomes more precise. 
\textit{Social Networks: 3, high impact}; Deepfakes and other modified text, voice, and videos are widely posted through social networks.

Let $S$ represent the total score for a suspected emerging DI, $n$ represent the total number of attributes considered, and $A_i$ represent the impact score for the $i$-th attribute, namely, low = 1, medium = 2, and high = 3 as defined above.
The mathematical formula for Model 2 is as follows:
\[
S = \sum_{i=1}^{n} A_i
\]
Then, a DI is flagged in our example if $S \geq T$ where $T$ is the threshold score (e.g.,  $T = 9$). For flagged DIs, we shall map their most prominent (highest scores) attributes to TTPs that historically correspond to those DI attributes. For example, Table \ref{Tab:mapping} contains references to TTPs for corresponding DIs. Notably, since 2007, DIs classified as having the prominent attribute of \textit{cyber communications pathways} map to \textit{content creation} and \textit{data manipulation} adversarial tactics \cite{mitre, DISARM}. Suspected DIs with matching prominence may require defense preparation similar to that of their historically related DIs. With this correlation, investigators can inform proactive defense strategies, enabling stakeholders to mitigate potential threats before large-scale deployment.

For the example in Table \ref{tab:model3}, the prominent attributes are \textit{Cyber Communications Pathways} and \textit{Social Networks}. We find historical DIs with prominent attribute similarities in Table \ref{Tab:mapping}. For example, \textit{social media} introduced social network accounts and expanded communications. Investigators shall benefit from researching successful defense strategies used with these related DIs. DISARM and MITRE ATT\&CK frameworks include ``blue team" defenses or countermeasures for cyber attack DIs useful for this predictive methodology. One such countermeasure listed on DISARM suggests ``inoculating populations through media literacy training" to counter related and relevant TTPs like information pollution and distorted facts \cite{DISARM}. Stakeholders may benefit from employing this defense against \textit{AI Synthetic Media}.

\subsubsection{Relationship Between Models 1 and 2}

Model 1 only predicts the year when the $(n+1)$th DI will emerge, but no information about its attributes. Model 2 aims to predict the attributes of the $(n+1)$th DI, which could guide the design of proactive defense. Collectively, we can use Model 1+2 to predict the emergence year of the $(n+1)$th DI and its attributes, leading to more information to guide the design of proactive defense.

\subsubsection{Validating Predictions}
\label{sec:robotic}

This section validates the prediction results in the yearly time intervals and shows the predicted DI with the associated and relevant attributes.\\

{\bf Experimental Evaluation: The 2021 ``Missing'' Data}

According to Model 1, another DI should have emerged after 2017, indicating that data is ``missing" for the next predicted time interval, $[2021, 2022]$. We consider this benchmark for real-world validation. Based on our methodology criterion and related research, we provide a prediction that shall be considered for our timeline: \textit{robotic humanoids}. This prediction may validate our methodology.

\begin{insight}
[2021 DI Prediction: Robotic Humanoids] Based on Model 1's calculation, a new DI should have emerged around 2021. Notably, the robotic humanoid ``Ameca'' was introduced in 2021 \cite{ameca}. 
\end{insight}

\begin{figure}[htbp!]
    \centering
    \includegraphics[width=0.7\linewidth]{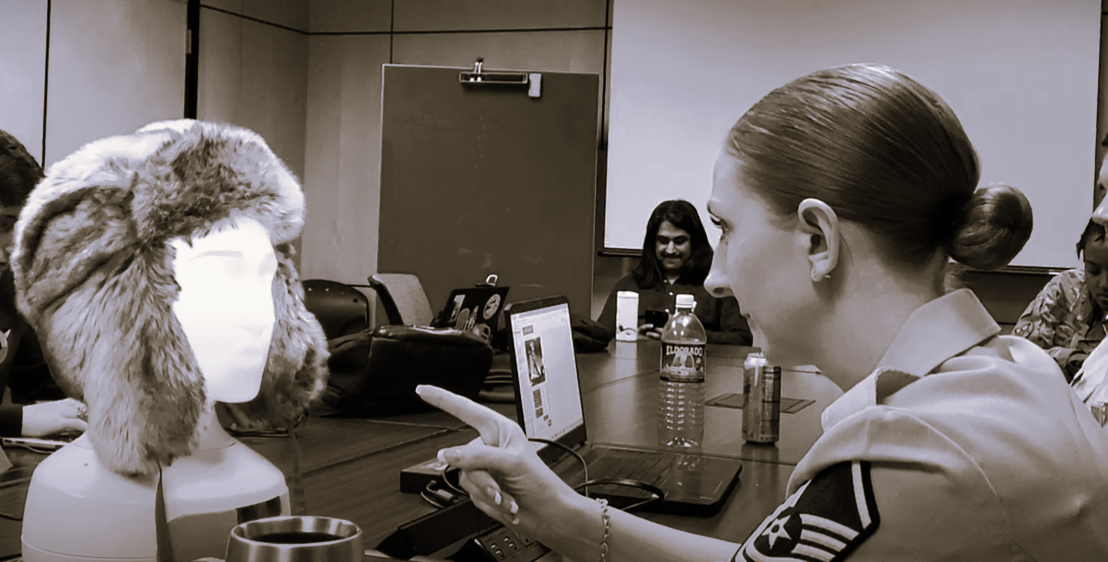}
    \caption{``Furhat" Robot experimentation at the US Air Force Academy (2023). Credit: Bonnie Rushing. The appearance of U.S. Department of Defense (DoD) visual information does not imply or constitute DoD endorsement.}
    \label{fig:furhat}
\end{figure}

The missing 2021 data point may be \textit{robotic humanoids.}
These human-shaped robot technologies have not been widely employed as weaponized conduits for cyber cognitive attacks. However, the employment of robotic humanoids for cognitive attacks seems probable based on our DI timeline, highlighting the weaponizing of similar dual-use computer technologies \cite{ameca}. 
We would add any corresponding TTPs and expert commentary from newer studies to the dataset. These robotic humanoids can be exploited by accessing user data through embedded microphones, cameras, and other user input. Victims can be targeted in advanced ways through social engineering based on their data and preferences, and selected messaging can be broadcast to users based on attackers' goals. 2023 research explored these attributes and TTPs, analyzing users' decision-making and trust in a robot's advice \cite{furhat}. However, this project involved a ``Furhat'' robot (Figure \ref{fig:furhat}) instead of a robotic humanoid. Based on their potential employment, we calculate the prominent attributes as follows: \textit{expanded communication pathways $(A = 3)$, expanded access to data collection $(A = 3)$, and advanced targeting $(A = 3)$}; $S = 9$; $T = 9$; $S \geq 9$; which means that it meets our DI inclusion threshold. Historic DIs with similar prominent attributes connected to TTPs include spearfishing, real-time communications, location tracking, and manipulation platform algorithms/exposure. These adversarial TTPs are connected to defensive TTPs within the DISARM and MITRE ATT\&CK frameworks, which will guide our proactive defense design in this paper.

This discovery may be validated if robotic humanoids are weaponized to deliver cognitive attacks. Cognitive-attacking robots may be in our future!

\subsubsection{Experimental Findings}
\label{sec:Analysis}

If our experiment's Model 1 anticipated time intervals continue, we can predict when subsequent cognitive attack DIs will emerge in future years. We highlight the mathematical outputs: 2021.6, 2026.2, 2030.8, 2035.4

These data start with our final data point from Table \ref{Tab:mapping}, year 2017. We calculated future years using the linear regression fitter formula $Y=-460.5 + 0.23X$, which can compute the subsequent DIs' years of emergence, $X$, by setting $Y_{12}$ and higher. We recorded the results in Model 1's column, showing that a new DI should emerge approximately every \textit{4.6 years}.

\subsection{Designing Proactive Defense}

Based on the experimental findings, our predictions may directly inform real-world cybersecurity strategies and proactive defenses. The timeline intervals from Model 1 and qualitative attributes from Model 2 may allow for early identification of threats. For example, one prediction from our experiment indicated that humanoid robots may be used maliciously in cognitive attacks. 
We calculated the prominent attributes for robotic humanoids: \textit{expanded communication pathways, expanded access to data collection, and advanced targeting}, using Model 2 to ensure the DI threshold was met. Further, we discovered that historic DIs with similar prominent attributes connected to TTPs such as spearfishing, real-time communications, location tracking, and manipulating platform algorithms/exposure. These adversarial TTPs are connected to defensive TTPs within the DISARM and MITRE ATT\&CK frameworks \cite{DISARM, mitre}, which we leverage to design proactive defense in this section.

\subsubsection{Defensive Strategy and Application}

Based on the predicted attributes of the DIs, we can leverage them to proactively design defenses by mapping adversarial and defensive TTPs to emerging DIs based on historical DIs with similar prominent attributes. This will enable investigators to predict specific attack details and related countermeasures from the DISARM and MITRE ATT\&CK frameworks. This methodology may be applied to any DI's adversarial TTPs, including our case study on robotic humanoids.

For our DI example of robotic humanoids, we discovered historic DIs with similar prominent attributes (e.g., algorithm targeting DI [introduced automated targeting], tracking and data collection DI [introduced automated data collection], and e-mail DI [introduced cyber communications]) connected to TTPs such as spearfishing, real-time communications, location tracking, and manipulate platform algorithms/exposure in Section \ref{sec:robotic}. These adversarial TTPs are connected to defensive TTPs within the DISARM and MITRE ATT\&CK frameworks, which we leverage to design proactive defense in this section.

Thus, we analyze defensive TTPs (if available) listed in the frameworks for associated adversarial TTPs. Specifically, robotic humanoid's prominent attributes connected to TTPs include spearfishing, real-time communications, location tracking, and manipulating platform algorithms/exposure. The data for robotic humanoid DI's prominent attributes, similar historic DIs, and related adversarial TTPs that map to defensive TTPs are listed in Table \ref{tab:defensemap}. 

\begin{table}[H]
   \centering
    \small 
    \renewcommand{\arraystretch}{1.1} 
    \setlength{\tabcolsep}{5pt} 
\begin{tabular}{|>{\raggedright\arraybackslash}p{0.21\linewidth}|>{\raggedright\arraybackslash}p{0.16\linewidth}|>{\raggedright\arraybackslash}p{0.21\linewidth}|>{\raggedright\arraybackslash}p{0.21\linewidth}|} \hline     
\textit{Prominent Attributes} & \textit{Similar DIs}& \textit{Adversary TTPs} &\textit{ Defensive TTPs} \\ \hline 
Communi-cation pathways& E-mail & spearfishing, real-time communications & MITRE ATT\&CK: User Training  \\ \hline 
Data Collection & Tacking \& Data Collection& location tracking & MITRE ATT\&CK: Enterprise policy \\ \hline 
Victim Targeting & Algorithm Targeting & manipulate platform algorithms \& exposure& DISARM: Change Search Algorithms for Disinformation Content \\ \hline

\end{tabular}
\caption{Defensive TTP Mapping for Humanoid Robots DI}
\label{tab:defensemap}
\end{table}

 These defensive TTPs within the DISARM and MITRE ATT\&CK frameworks describe how each countermeasure can be applied to mitigate attack risk. For our examples in Table \ref{tab:defensemap}, the following information can be found in the frameworks' defensive TTPs.

\textit{ MITRE ATT\&CK: User Training.} ``Train users to be aware of access or manipulation attempts by an adversary to reduce the risk of successful spearphishing, social engineering, and other techniques that involve user interaction. \cite{mitre}"

\textit{ MITRE ATT\&CK: Enterprise policy.} ``If devices are enrolled using Apple User Enrollment or using a profile owner enrollment mode for Android, device controls prevent the enterprise from accessing the device’s physical location. This is typically used for a Bring Your Own Device (BYOD) deployment. \cite{mitre}"

\textit{ DISARM: Change Search Algorithms for Disinformation Content.} ``Includes `change image search algorithms for hate groups and extremists' and `Change search algorithms for hate and extremist queries to show content sympathetic to opposite side' \cite{DISARM}."

Stakeholders can benefit by employing these mitigation strategies specifically designed to counter related adversarial TTPs before the DI's large-scale implementation. For example, with MITRE ATT\&CK: \textit{Enterprise policy} (mitigation ID: M1012) TTP, interdisciplinary leaders should proactively implement policy and controls for devices and users' location and tracking for security before humanoid robots become widely available. The MITRE ATT\&CK framework contains extensive examples of incidents and specific defense descriptions \cite{mitre}. These actionable defense strategies leverage our predictive models to inform cybersecurity, policymaking, and public awareness. Through cross-disciplinary collaboration and continuous refinement, the models can help anticipate and mitigate the impacts of emerging DI in cyber cognitive attacks.

\section{Discussion}
\label{sec:discussion}

This methodology aims to solve complex problems for emerging DIs. These models can advance with AI and machine learning, growing in scalability with larger datasets. The novel model integration with frameworks for DI prediction benefits global societies. This methodology can be employed responsibly at scale through increasingly diverse inputs and refinement and by incorporating machine learning techniques such as time series forecasting neural networks when the dataset grows more extensive in the future, enhancing the framework. 

This framework may be adapted for other studies and domains. Investigators may be interested in predictive analysis of TTPs for other topics in IT,  computer science, government, and business services. MITRE ATT\&CK contains TTPs relating to interdisciplinary domains, allowing researchers to adapt our methodology for diverse studies at larger scales, far beyond cyber cognitive attack DIs \cite{mitre}.  For example, future studies may use AI-driven web scraping or natural language processing (NLP) to analyze large datasets, such as scientific publications, patents, or news articles, to identify early indicators of threats. Tools like GPT-based models, semantic search, or AI-powered data extraction systems can improve data collection and filtering in the future. AI may also be able to process and analyze datasets (e.g., technological trends, funding announcements, R\&D reports) to uncover hidden patterns that signal upcoming threats or DIs.

\noindent{\bf Limitations}.
First, our methodology focuses on macro-group DIs, which are or refer to our DI definition. It is interesting to adopt or adapt the methodology to incorporate micro-group DIs, which refer to the subsets of DI upgrades that do not meet the DI definition. This would lead to a more comprehensive understanding of the DI landscape. Furthermore, our methodology focuses on the malicious use of DIs without considering DIs for defending against cognitive attacks. It is interesting to extend our methodology to accommodate this perspective. Next, model 1 may assume a single DI emergence per year, which is a limitation of these calculations. Finally, the scoring methodology in Model 2 may be considered subjective based on the investigators' understanding of DIs.

Second, our case study also has some limitations, as follows. Our perception of DI uniqueness, importance, and usage is subjective, making the resulting dataset subjective. It is an interesting open problem to investigate methods to reduce subjectivity, for instance, by leveraging a team of researchers with relevant but diverse expertise.

\noindent{\bf Risks and Mitigation}.
The study has three potential risks. First, our methodology relies on the completeness of historical data, meaning that our case study may have missed some relevant DIs. To mitigate the risk, one can apply our model to various choices of datasets.
Second, stakeholders may resist extra training events or spending scarce resources on awareness campaigns or research. To mitigate this risk, leaders should highlight cost-benefit analysis and provide scalable implementation options.
Third, malicious actors may abuse the study's results. The mitigation includes publicizing insights for worldwide researchers without providing explicit procedures for conducting adversarial actions.

\section{Conclusion}
\label{sec:Conclusion}
This study introduces a novel methodology for predicting the emergence of DIs that may be exploited in cyber cognitive attacks, and offers actionable strategies for proactive defense. By analyzing historical trends and leveraging predictive models, such as the Linear Regression and Qualitative Attribute models, we provide a framework to forecast both the timing and attributes of future DIs. This enables the cybersecurity community to anticipate threats and allocate resources effectively for defense.

Our findings highlight the increasing sophistication of cognitive attacks enabled by emerging technologies like AI, social media, and humanoid robotics. The predictive models not only forecast critical advancements, such as the emergence of humanoid robots for cognitive attacks in 2021, but also underline the importance of integrating collaborative defenses. These contributions bridge predictive analytics with frameworks like MITRE ATT\&CK and DISARM to map adversarial TTPs to appropriate countermeasures. Future work should validate and extend these models with larger datasets and incorporate evolving technologies to refine predictions. Additionally, collaboration across disciplines will be essential to mitigate the societal impacts of these innovations and ensure resilience against adversarial advancements.

By combining historical analysis with predictive insights, this research provides a foundation for proactive strategies to mitigate the disruptive impacts and risks of cyber cognitive attacks. With continuous refinement and broader validation, this methodology can empower stakeholders to stay ahead of emerging threats and protect individuals, organizations, and societies. This methodology can empower populations by addressing cyber challenges such as disinformation and sustaining overall progress in democratic and socioeconomic development through proactive defenses. 

\paragraph{Data availability statement}
 The authors confirm that the data supporting the findings of this study are available within the article and/or its supplementary materials.

\newpage
\printbibliography

@article{breese1998empirical,
    title = {Empirical Analysis of Predictive Algorithms for Collaborative Filtering},
    author = {Breese, John S. and Heckerman, David and Kadie, Carl},
    journal = {Proceedings of the 14th Conference on Uncertainty in Artificial Intelligence (UAI-98)},
    year = {1998},
    url = {https://arxiv.org/pdf/1301.7363},
    urldate = {2024-09-30}
}

@article{Crupi_Mejova_Tizzani_Paolotti_Panisson_2022, title={Echoes through Time: Evolution of the Italian COVID-19 Vaccination Debate}, volume={16}, url={https://ojs.aaai.org/index.php/ICWSM/article/view/19276}, DOI={10.1609/icwsm.v16i1.19276}, number={1}, journal={Proceedings of the International AAAI Conference on Web and Social Media}, author={Crupi, Giuseppe and Mejova, Yelena and Tizzani, Michele and Paolotti, Daniela and Panisson, André}, year={2022}, month={May}, pages={102-113} }

@article{lin2023mapping,
  title={Mapping language literacy at scale: a case study on Facebook},
  author={Lin, Yu-Ru and Wu, Si and Mason, Winter},
  journal={EPJ Data Science},
  volume={12},
  number={1},
  pages={13},
  year={2023},
  publisher={Springer},
  doi={10.1140/epjds/s13688-023-00388-4},
  url={https://doi.org/10.1140/epjds/s13688-023-00388-4}
}

@article{longtchi2024internet,
  title={Internet-Based Social Engineering Psychology, Attacks, and Defenses: A Survey},
  author={Longtchi, Theodore Tangie and Rodriguez, Rosana Montanez and Al-Shawaf, Laith and Atyabi, Adham and Xu, Shouhuai},
  journal={Proceedings of  IEEE},
volume={112},
number={3},
pages={210-246},
  year={2024},
}

@article{DISARM,
    author = {DISARM},
    title = {DISARM Framework Explorer},
    year = {2024},
    note = {https://disarmframework.herokuapp.com/tactic/10/view
}}

@misc{mitre,
  author       = {{MITRE Corporation}},
  title        = {MITRE ATT\&CK Framework},
  year         = 2024,
  url          = {https://attack.mitre.org/},
  note         = {Accessed: 2024-12-21}
}

@report{elon,
  title        = {The Future of Free Speech, Trolls, Anonymity, and Fake News Online},
  author       = {Rainie, Lee and Anderson, Janna and Albright, Jonathan},
  institution  = {Pew Research Center},
  year         = 2017,
  month        = mar,
  day          = 29,
  url          = {https://www.pewresearch.org/internet/2017/03/29/the-future-of-free-speech-trolls-anonymity-and-fake-news-online/}
}

@article{ANTONIO2023122274,
title = {Contextual factors of disruptive innovation: A systematic review and framework},
journal = {Technological Forecasting and Social Change},
volume = {188},
pages = {122274},
year = {2023},
issn = {0040-1625},
doi = {https://doi.org/10.1016/j.techfore.2022.122274},
url = {https://www.sciencedirect.com/science/article/pii/S0040162522007958},
author = {Jerome L. Antonio and Dominik K. Kanbach}
}

@report{CSET,
    title = {AI and the Future of Disinformation Campaigns},
    author = {{Center for Security and Emerging Technology (CSET)}},
    institution = {Georgetown University},
    year = {2020},
    type = {Report},
    url = {https://cset.georgetown.edu/publication/ai-and-the-future-of-disinformation-campaigns/},
    urldate = {2024-09-27}
}

@article{ieuniversity,
    author = {Irene Pujol Chica and Quynh Dinh Da Xuan},
    title = {THE BATTLE FOR THE MIND, Understanding and addressing cognitive
warfare and its enabling technologies},
journal = {ie university},
    year = {2024},
url = {https://static.ie.edu/CGC/CGC_TheBattleofTheMind_2024.pdf},
}

@article{NATOscience,
    author = {Bernard Claverie and François du Cluzel},
    title = {“Cognitive Warfare”: The Advent of the Concept of “Cognitics” in the Field of Warfare},
journal = {Cognitive Warfare, First NATO scientific meeting on Cognitive Warfare},
    year = {2021},
url = {https://www.researchgate.net/publication/359991886_Cognitive_Warfare_The_Advent_of_the_Concept_of_Cognitics_in_the_Field_of_Warfare},
}

@article{blacksea,
    author = {Olga R. Chiriac},
    title = {COGNITIVE WARFARE IN THE 21ST CENTURY GREAT POWER COMPETITION
– FRAMING OF MILITARY ACTIVITY IN THE BLACK SEA},
    year = {2021}
}

@book{Linan,
    author = {Linan Huang and Quanyan Zhu},
    title = {
Cognitive Security: A System-Scientific Approach},
    publisher = {Springer},
    year = {2023}
}

@article{ox,
    author = {William Hersch and Melissa McLain},
    title = {Inside the Gates: Cultivating Cognitive Security to Defend the Homeland},
journal={Journal of Indo-Pacific Affairs},
    year = {2024}
}

@misc{motherboard2010disinformation,
  author       = {Motherboard},
  title        = {Disinformation in the Age of the Internet},
  year         = {2010},
  howpublished = {\url{https://web.archive.org/web/20120324004022/}},
  note         = {Archived version retrieved from the Wayback Machine},
  urldate      = {2024-12-25}
}

@article{email,
    author = {Sarah Left},
    title = {Email timeline
},
    year = {2002}
}

@article{bots,
    author = {Noémi Bontridder and Yves Poullet},
    title = {The role of artificial intelligence in disinformation
},
journal={Data and Policy},volume={3}, number={3}, pages={1-21},
    year = {2021}
}

@article{trolls,
    author = {Jessikka Aro},
    title = {The Cyberspace War: Propaganda and Trolling as Warfare Tools},
journal={European View}, volume={15}, number={1}, pages={121-132},
    year = {2016}
}

@article{apps,
    author = {Christine Pitt and Jeannette Paschen and Jan Kietzmann and Leyland F. Pitt and Erol Pala},
    title = {Artificial Intelligence, Marketing, and the History of Technology: Kranzberg’s Laws as a Conceptual Lens},
journal={Australasian Marketing Journal}, volume={31}, number={1}, pages={1-89},
    year = {2021}
}

@article{socialmedia,
    author = {K Sajithra and Rajindra Patil},
    title = {Social Media – History and Components},journal={IOSR Journal of Business and Management (IOSR-JBM)}, volume={7}, number={1}, pages={69-74},
    year = {2013}
}

@article{mud,
    author = {F Sudweeks and H Hrachovec and C Ess},
    title = {Flow Experience and Interaction in Online Gaming},
    year = {2008}, journal={Proceedings Cultural Attitudes Towards
Communication and Technology}, volume={1}, number={1}, pages={410-421}
}

@inproceedings{trackers,
author = {Englehardt, Steven and Reisman, Dillon and Eubank, Christian and Zimmerman, Peter and Mayer, Jonathan and Narayanan, Arvind and Felten, Edward W.},
title = {Cookies That Give You Away: The Surveillance Implications of Web Tracking},
year = {2015},
isbn = {9781450334693},
publisher = {International World Wide Web Conferences Steering Committee},
address = {Republic and Canton of Geneva, CHE},
url = {https://doi.org/10.1145/2736277.2741679},
doi = {10.1145/2736277.2741679},
booktitle = {Proceedings of the 24th International Conference on World Wide Web},
pages = {289–299},
numpages = {11},
location = {Florence, Italy},
series = {WWW '15}
}

@article{trackers2,
    author = {Shane Morris and David Gurzick and Sean Guillory and Glenn Borsky},
    title = {Countering Cognitive Warfare in the Digital Age: A Comprehensive Strategy for Safeguarding Democracy against Disinformation Campaigns on the TikTok Social Media Platform},
    year = {2024}
}

@INPROCEEDINGS{iot,
  author={Suresh, P. and Daniel, J. Vijay and Parthasarathy, V. and Aswathy, R. H.},
  booktitle={2014 International Conference on Science Engineering and Management Research (ICSEMR)}, 
  title={A state of the art review on the Internet of Things (IoT) history, technology and fields of deployment}, 
  year={2014},
  volume={},
  number={},
  pages={1-8},
  doi={10.1109/ICSEMR.2014.7043637}}

@incollection{VR,
title = {Virtual Reality: Definitions, History and Applications},
editor = {RA Earnshaw and MA Gigante and H Jones},
booktitle = {Virtual Reality Systems},
publisher = {Academic Press},
address = {Boston},
pages = {3-14},
year = {1993},
isbn = {978-0-12-227748-1},
doi = {https://doi.org/10.1016/B978-0-12-227748-1.50009-3},
url = {https://www.sciencedirect.com/science/article/pii/B9780122277481500093},
author = {Michael A. Gigante}
}

@article{ameca,
    author = {Celia Nieto Agraz and Pascal Hinrichs and Marco Eichelberg and Andreas Hein},
    title = {Is the Robot Spying on me? A Study on Perceived Privacy in Telepresence Scenarios in a Care Setting with Mobile and Humanoid Robots},
journal={International Journal of Social Robotics},
    year = {2024}
}

@article{furhat,
    author = {Jonah Nascimento and Niyah Martinez and Harley Anez and Spenser Schwalm and Kekoa Gross and Benjamin Pederson and Amiyah Breeding and Ewart de Visser and Steven Hadfield and Ali Momen and Melissa Mclain and Bonnie Rushing and Christopher Gausepohl and Chad Tossell},
    title = {Robot Moral Advice in a High Stakes Military Environment},
journal={6th International Conference on Intelligent Human Systems Integration: Integrating People and Intelligent Systems},
    year = {2023}
}

@String{IEEE =   "IEEE"}
\end{document}